# All-optical directional switching of non-thermal photocurrents in plasmonic nanocircuits

Roméo Zapata[1°], Diana Singh[2°], Obren Markovic[1], Chantal Hareau[1], Xingyu Yang[1], Ye Mou[3], Catherine Schwob[1], Bruno Gallas[1], Maria Sanz-Paz[1], Gérard Colas-des-Francs[2], Alexandre Bouhelier[2] and Mathieu Mivelle[1*]

[1]Sorbonne Université, CNRS, Institut des NanoSciences de Paris, INSP, F-75005 Paris, France

[2]Université Bourgogne Europe, CNRS, Laboratoire Interdisciplinaire Carnot de Bourgogne ICB UMR 6303, 21000 Dijon, France

[3]School of Electronic and Information Engineering, Ningbo University of Technology, No. 201, Fenghua Road, Jiangbei District, Ningbo, Zhejiang, China

° Equal contribution

*Corresponding authors:

mathieu.mivelle@sorbonne-universite.fr

**Abstract**

Controlling the flow of electricity in metallic circuits with light is a key goal for future optoelectronics. In this work, we demonstrate all-optical generation and directional control of non-thermal drift photocurrents in a plasmonic gold wire. We attribute this phenomenon to the Inverse Faraday Effect and show that the current's direction can be precisely reversed at a subwavelength scale by tailoring the incident light's polarization or laser beam position. A bespoke polarization modulation technique is employed to unambiguously separate ultrafast drift currents from the ubiquitous photothermal background. We further reveal a collaborative mechanism where macroscopic thermal gradients, acting as a driving force, are used to extract and remotely detect the locally-generated nanoscale photocurrents. This robust control and detection scheme paves the way for reconfigurable, all-optical nanocircuitry capable of ultrafast on-chip processing.

**Keywords:** Photocurrent, Plasmonics, Inverse Faraday effect, Nanophotonics, Optoelectronics, Nonlinear Optics

**Introduction**

A photocurrent is the flow of electric charges generated when photons are absorbed by a material. This phenomenon is foundational to numerous groundbreaking technologies, including light detection, solar photovoltaics, lightwave electronics, and THz spectroscopy and imaging[1]. For physicists, a direct photocurrent is also a unique probe for investigating key processes across vast spatio-temporal scales, from femtoseconds to milliseconds and from the microscale to the mesoscale. As a result, photocurrents have been instrumental in advancing our understanding of low-dimensional structures[2], topological materials[3], and systems exhibiting symmetry breaking[4,5], offering a window into the behavior of out-of-equilibrium transient states[6,7].

The conversion of light into an electrical signal can stem from a wide collection of physical mechanisms. In all-metal plasmonic circuits, photocurrents can be generated by different contributing mechanisms a significant debate exists between two primary categories of effects. Foremost is the photothermoelectric (PTE) effect, where optical absorption leads to localized heating. The resulting temperature gradients ($\nabla T$), combined with spatial variations in the material's Seebeck coefficient ($S$), produce a current in a closed-circuit configuration[8-15]. Other contributors to photocurrent generation include various nonthermal phenomena[16] such as photon drag (momentum transfer from photons to electrons)[17,18], hot-electron injection[19], and other plasmoelectric effects[20,21]. However, significant confusion often remains regarding the correct assignment of these mechanisms in experimental observations, highlighting the complex interplay between a material's optical and electronic properties. Despite this complexity, all-metal plasmonic systems present a compelling platform to study, as they allow for the fine-tuning of local optical fields and the generation of electric currents on the same material, thereby avoiding the constraints of complex hetero-structures.

Despite recent efforts, a comprehensive framework for precisely manipulating direct photocurrents at the nanoscale using plasmonics is still largely missing. A robust strategy to engineer their spatial distribution, control their amplitude, and, most importantly, dictate their direction has yet to be fully established. Here, we explore a physical process that has often been overlooked for photocurrent generation in metal nanostructures: nonlinear optical forces. This occurs when a ponderomotive force induces a drift current within the skin depth of a metal. This mechanism for photocurrent generation offers several advantages: it is occurring on a timescale of a few femtoseconds[22], it persists only for the duration of the light exposure, and it can be precisely manipulated by tuning the optical field's intensity, gradients, or polarization.

A drift current can be understood as a form of optical rectification and is linked to the Inverse Faraday Effect (IFE) in metals[23], a nonlinear process that promotes the magnetization of a material through purely optical excitation[22,24-30]. While traditionally associated with circularly polarized light in symmetric structures, recent works have shown that plasmonic nanostructures unveil new facets of the IFE, enabling its generation with linear polarizations[31], inducing chiral responses[32,33], or even reversing its symmetry[34]. Building on these, we demonstrate that such nonlinear drift currents provide a powerful and direct pathway to generate photocurrents. Their direction is deterministically controlled by the polarization of the incident light[31], thus offering a new degree of freedom for designing advanced optoelectronic devices[32,33].

Herein, we show the generation at the nanoscale of drift photocurrents $J_d$ generated by IFE (Eq. 1) within a wire-bonded gold nano strip by placing plasmonic nanostructures in its near-field. In the framework of IFE, the drift current is written[23,24] as

$$\boldsymbol{J_d} = \frac{1}{2en} Re\left(\left(-\frac{\nabla \cdot (\sigma_\omega \boldsymbol{E})}{i\omega}\right) \cdot (\sigma_\omega \boldsymbol{E})^*\right) \quad (1)$$

Where $e$ is the electron charge, $n$ is the equilibrium electron density, $\sigma_\omega$ is the dynamic conductivity, and $\boldsymbol{E}$ represents the optical electric field.

We found that the direction of these photocurrents is fully controllable at the subwavelength scale by tailoring the polarization of the incident light or the position of the laser beam. Furthermore, by selectively modulating either the intensity or the polarization of the optical excitation, we successfully disentangle the drift contribution from the underlying photothermal response. This approach unequivocally demonstrates that the drift current's direction is polarization-dependent and, remarkably, temperature-independent. Finally, we reveal a collaborative effect between the drift currents and the metal's thermal response. We show that local thermal gradients can be harnessed to probe and transduce nanoscale drift photocurrents, enabling their remote collection by our detection. This polarization-based control over the generation and direction of photocurrents in plasmonic metals paves the way for all-optical nanocircuitry for on-chip signal processing and information transduction or reconfigurable optoelectronic logic elements. Such a scheme promises ultrafast on-chip processing, as it harnesses the quasi-instantaneous nature of the underlying Inverse Faraday Effect.

**Results**

Firstly, to disentangle the drift photocurrents from their thermal counterparts, we performed experiments by exciting a bare gold strip without nanostructures (Fig. 1) with a 780 nm laser focused through a microscope objective (NA=1.49) and scanned along the structure in the XY plane. The generated photocurrents at each laser position, regardless of their origin, were measured using a lock-in detection scheme (see Fig. S1 for the full experimental setup). Two distinct laser modulation schemes were employed: an intensity modulation for with a static linear polarization of +45° or -45° (Fig. 1a) and a polarization modulation, switching between +45° and -45° relative to the strip's axis (Fig. 1b). This strategy serves a dual purpose: it enhances the photocurrents signal-to-noise ratio via the lock-in amplifier detection and, more importantly, it helps to isolate the different mechanisms contributing to the current signal. Indeed, since the optical absorption of the structure is identical for both ±45° linear polarizations, any resulting thermal effect constitutes a DC background signal that is rejected by the lock-in amplifier during the polarization modulation. In contrast, intensity modulation contains both thermal and non-thermal photocurrent components, with the thermal contribution typically dominating. Therefore, polarization modulation effectively isolates the non-thermal, polarization-dependent component of the photocurrent.

Figs. 1c and d show the photocurrent maps reconstructed by raster scanning the sample through the laser focus. The signal is extracted by the lock-in amplifier under an intensity modulation and for two different orientations of the incoming polarization (±45° with respect to the strip) and bias voltage set at -75 mV. As expected for a predominantly thermal response, wherein laser-induced thermal gradients generate a thermoelectric (Seebeck) current driven by the bias 10. Although not shown here, we verified that the direction of the current flow switches with the voltage polarity. Importantly, the detected photocurrent remains constant across the entire length of the nanowire, and is independent of the polarization orientation (see Fig. S2 for conductivity measurements and Fig. S3 for theoretical predictions of the photothermal response). Both +45° and -45 give the same current magnitude and distribution. In stark contrast, a polarization modulation (Fig.1e) reveals a fundamentally different behavior. A transverse profile across the gold strip (Fig. 1h) shows that the photocurrent polarity, represented in Fig. 1e by the red and blue colors superimposed on their intensity, inverts, exhibiting a characteristic negative-positive-negative signature, which is inconsistent with a purely thermal origin.

However, this map does not directly reflect the polarization-dependent current distribution directly. The photoelastic modulator (PEM) used for polarization modulation introduces a known parasitic positional modulation of the laser beam[35], resulting in a dual modulation of both polarization and position. Consequently, the demodulated signal corresponds not to the simple difference in currents between the two polarization states, but rather to the spatial derivative of the difference[36,37] (see SI for complete demodulation description). Therefore, for a rigorous comparison between experiment and theory, we calculated the spatial derivative (Fig. 1f) of the theoretical drift current difference (Fig. 1g), calculated through Eq. 1.

An excellent agreement between theory and experiment is confirmed by comparing the photocurrent maps in Figs. 1e and 1f and their corresponding transverse line scans (Fig. 1h). Furthermore, by numerically integrating the experimental signal from Fig.1h (orange dotted line in Fig. 1i), we retrieve the theoretical drift current's profile (black solid line in Fig. 1i).

Two key conclusions can be drawn from these initial experiments. First, as evidenced by the polarity and amplitudes of the measured photocurrent, the modulation of the polarization is an effective approach to isolate non-thermal photocurrents by suppressing the photothermal background. Second, our results demonstrate that the direction of drift photocurrents can be precisely controlled and reversed at the nanoscale within a simple metallic wire by tuning external optical parameters, such as the laser's polarization and position with respect to the nanowire.

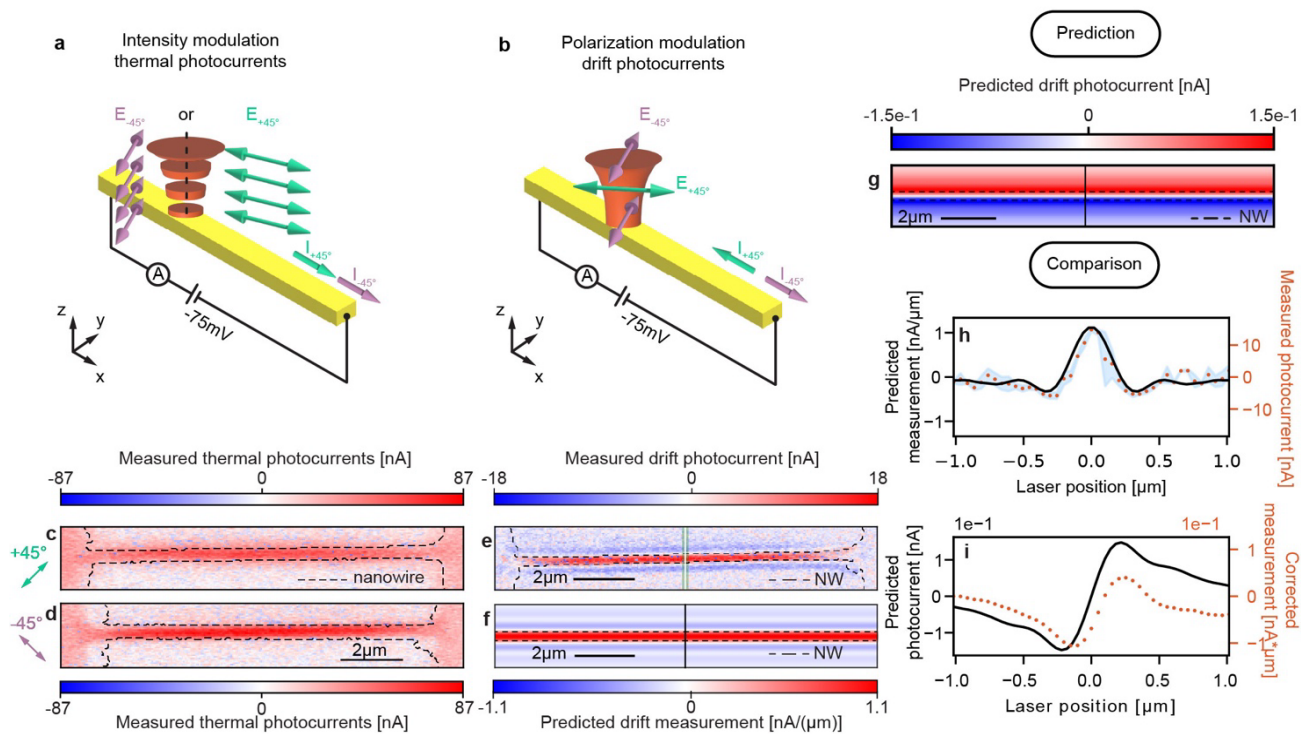

**Figure 1. Isolating and controlling polarization-dependent drift photocurrents on a bare gold wire. a, b,** Schematics illustrating the experimental setup for intensity and polarization modulation (between +45° and -45°), respectively. A bias voltage of -75 mV was applied for all measurements. **c, d,** Photocurrent maps under intensity modulation for static linear polarizations of +45° (**c**) and -45° (**d**), showing a constant polarity, characteristic of a dominant photothermal response. Dashed lines represent the location of the gold wire as extracted from the confocal reflection map (see details in Fig. S2, together with conductivity measurements). **e,** The map recorded using polarization modulation isolates the non-thermal signal, revealing a clear sign inversion across the wire's width, a signature of a polarization-dependent drift current. **f,** The corresponding theoretical map, representing the spatial derivative of the intrinsic theoretical drift current distribution in **g**, (based on the Inverse Faraday Effect) to account for an experimental artifact and provide a direct counterpart to (**e**). **h,** Transverse line scans comparing the experimental cross-sections from the green region in (**e**) (orange points) and the theoretical derivative from (**f**) (black line), showing excellent agreement (the blue shaded region represents the standard deviation of the averaged lines). **i,** Final comparison confirming that the numerically integrated experimental cross-section (orange dotted line) faithfully reproduces the theoretical profile (black solid line, taken from **g**).

Building on these initial findings, we moved to investigating a gold strip decorated with gold nanorods to locally tailor the electromagnetic fields and, consequently, the associated drift currents (Fig. 2). The nanorods were positioned along both sides of the strip, either in direct contact or separated by a nominal 60 nm gap. For this experiment, only polarization modulation was employed. As established previously, this modulation scheme ensures that thermal effects, which are identical for ±45° polarizations, are suppressed from the measurement. The resulting photocurrent map, recorded with a representative applied bias voltage of -50 mV, is shown in Fig.2a (see results with other bias voltages in Fig. S4). The map recovers the transverse negative-positive-negative pattern observed on the bare wire. However, two distinct new features emerge longitudinally. First, periodic discontinuities appear, predominantly along the top edge of the strip, which spatially correlate with the locations of the touching nanorods as confirmed by the SEM image (Fig. 2b). Second, longitudinal oscillations are now present in the collected signal across the top, central, and bottom parts of the wire in the horizontal direction, a signature not present in the bare wire case. We attribute these oscillations to interferences between forward and backward propagating plasmon waves created and reflected by the nanostructures along the horizontal direction of the strip.

A comparison with our theoretical model validates these observations. The photocurrent map in Fig. 2c, a magnified view of the region marked by a rectangle, is compared to the spatial derivative (Fig. 2d) of the theoretical current distribution (Fig. 2e). This derivative is calculated along the primary axis of the estimated PEM-induced beam wobble (indicated by the arrow in Fig. 2d). An excellent agreement is found between the experimental (Fig. 2c) and theoretical maps (Fig. 2d). Minor discrepancies are attributed to a simplification in the model, which treats the elliptical PEM-induced beam motion as linear. Notably, the simulation successfully reproduces the longitudinal oscillations arising from plasmon wave interference, as well as the faint, off-strip signals (dashed circles) resulting from the beam's spatial modulation, thereby underscoring the robustness of our results. This confirms again both the observation of drift currents and their successful manipulation via external control of the laser's polarization and position (see SI, Fig. S5, for the asymmetric wire decorated with nanorods on only one side, which also shows a conclusive agreement with theory).

Finally, a key observation is the abruptness of the polarity reversals of the photocurrents. These sign changes occur over a span of just 375 nm, corresponding to two pixels of the piezoelectric stage's scan. This demonstrates that the current's direction is controlled at a deeply subwavelength scale. Furthermore, while the discontinuity effect observed in Fig. 2 is most pronounced for the rods in direct contact (as the non-contacting ones in this specific sample had too large a gap), we confirm that the effect is also remarkable for closely-coupled, non-contacting rods, which produce similarly abrupt, subwavelength reversals (see Fig. S5 for a dedicated study on this configuration).

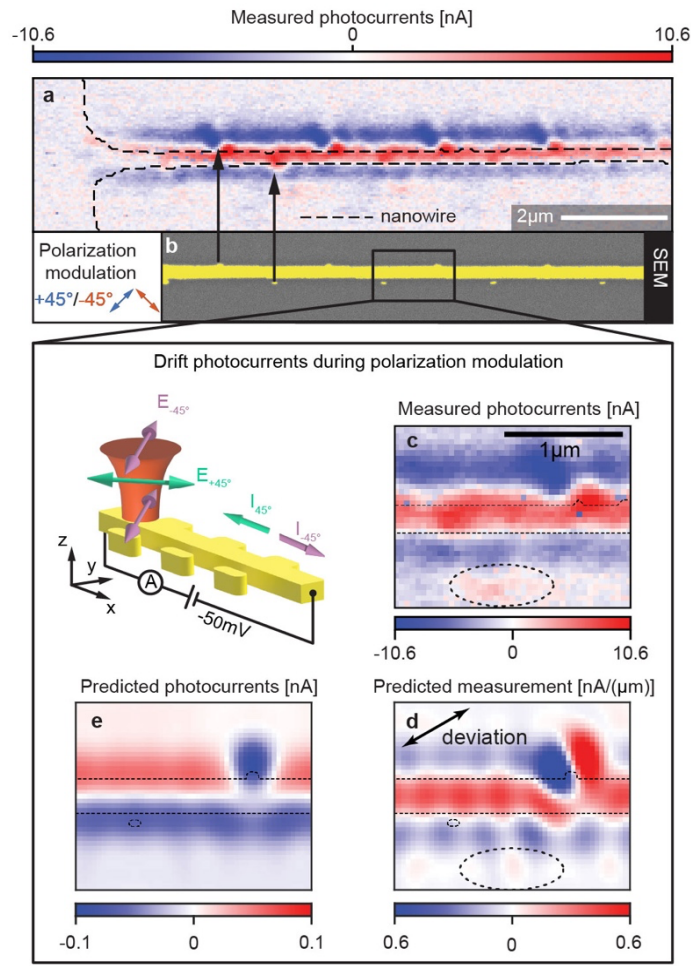

**Figure 2. Local and subwavelength control of drift photocurrents using plasmonic nanorods. a,** Experimental photocurrent map of the nanorod-decorated gold strip, recorded under polarization modulation. The map shows the transverse sign inversion (as in Fig. 1) overlaid with new periodic longitudinal discontinuities and oscillations. Dashed lines represent the location of the gold wire as extracted from the confocal reflection map (see details in Fig. S2). **b,** Scanning electron microscope (SEM) image of the sample, showing nanorods positioned alongside the strip, either in direct contact (top) or with a nominal 60 nm gap (bottom). These nanorod positions correspond to the discontinuities observed in (**a**). **c,** Magnified view of the experimental photocurrent map. **d,** The corresponding theoretical map, calculated as the spatial derivative of the intrinsic drift current distribution (**e**) to account for the experimental setup. **e,** Theoretical drift current distribution based on the Inverse Faraday Effect. Good agreement is found between the experimental map (**c**) and its theoretical counterpart (**d**). Dotted lines represent the location of the gold wire and the nanorods, while the dashed circles highlight faint, non-noise signals far from the strip, which are well reproduced in the simulation. The abrupt polarity reversals, occurring over just a few tens of nanometers, demonstrate deeply subwavelength control. A bias voltage of -50 mV was applied.

Finally, to better understand the interplay and collaboration between thermal gradients and drift photocurrents, a concluding experiment was conducted on this nanorod-decorated wire, with no applied bias voltage. Four distinct excitation schemes were investigated, as summarized in Fig. 3. The first two cases involved intensity modulation with a fixed linear polarization, oriented either perpendicular to the wire (Fig. 3a) or at 45° with respect to its axis (Fig. 3b). The latter two cases utilized polarization modulation, switching first between collinear and perpendicular polarizations (Fig. 3c), and subsequently between +45° and -45° polarizations (Fig. 3d).

As seen in Fig. 3, photocurrents are detected under all four excitation schemes, but only when the laser scans the distal ends of the wire near the electrodes. The current polarity is opposite at each end, a signature of the opposing thermal gradients between the hot wire and the cooler electrodes. These gradients act as a driving force, propelling the locally generated charge carriers in the circuit towards the current detector. At the center of the nanowire, these gradients effectively cancel each other out, and consequently, no net current is propelled towards the measurement system (see Fig.S3 for theoretical thermal gradient predictions).10

A closer analysis of the different modulation schemes reveals the distinct origins of the signal. Under intensity modulation (Figs. 3a and b), the thermal photocurrent distributions are qualitatively similar, though the signal is weaker for the 45° polarization due to a less efficient optical absorption (Fig. S6). Interestingly, a polarization modulated between 0° and 90° yields a nearly identical, thermally-dominated signal. This seemingly counterintuitive result stems from the fact that the near-field light distribution—thus the absorption—for these two orthogonal polarizations is different (Fig. S6) , like in the case of Figs. 3a and b. Consequently, the heat generation is modulated, and the lock-in technique fails to suppress the thermal background, which predominates the signal. In stark contrast, a polarization modulation between ±45° recovers the purely non-thermal drift current signature observed previously (Fig. 2a), but only at the wire ends where the thermal gradients are present to enable charge extraction, a condition that applies to all maps in Fig. 3. These results are significant for several reasons. First, they definitively confirm that ±45° modulation successfully isolates the polarization-dependent drift component from the thermal background. Second, they reveal a collaborative interplay between thermal and drift phenomena: the drift currents are generated by the local optical fields, while the macroscopic thermal gradients are required to "extract" and drive these currents to the detector. Finally, this synergic effect demonstrates a viable method for electrically-detecting nanoscale optical signals, opening exciting prospects for the development of all-optical nanocircuitry.

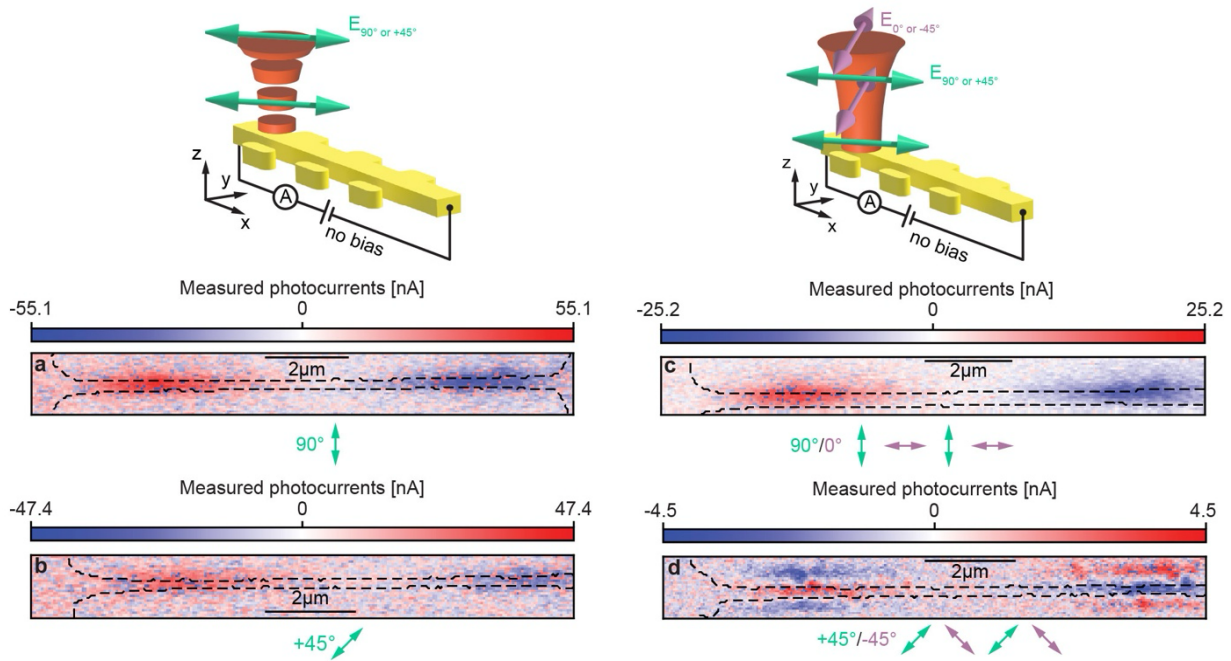

**Figure 3. Synergistic interplay of thermal gradients and drift currents without bias voltage.** Photocurrent generation on the nanorod-decorated wire with zero applied bias, using four excitation schemes. Photocurrent maps recorded under intensity modulation with polarization perpendicular (**a**) or at 45° (**b**) to the wire show a similar, thermally-driven signal confined to the wire ends, where macroscopic thermal gradients exist. The map recorded under polarization modulation between 0° and 90° (**c**) is also dominated by thermal effects, as the different near-field absorption for these polarizations prevents thermal background cancellation. In contrast, polarization modulation between ±45° (**d**) successfully isolates the non-thermal drift current signature, but *only* at the wire ends. This demonstrates a collaborative effect: the drift currents are generated optically by the IFE, while their remote detection is enabled by the driving force of the thermal gradients.

**Conclusion**

In conclusion, we have demonstrated the all-optical generation and deterministic directional control of non-thermal drift photocurrents in plasmonic gold circuits. By employing a specific polarization modulation scheme (±45°), we successfully isolated these drift currents, which originate from the Inverse Faraday Effect, from confounding photothermal backgrounds. Our experiments reveal that the current's direction can be precisely manipulated by the incident light's polarization and position, with plasmonic nanostructures enabling local control at a subwavelength scale. Furthermore, we uncovered a synergistic interplay between optical and thermal phenomena. While the drift currents are generated by a purely non-thermal optical mechanism, we showed that macroscopic thermal gradients can be harnessed as an effective means to "extract" and drive these nanoscale currents to a remote detector. This collaborative effect provides a robust method for probing local, optically-induced electronic phenomena from tens of centimeters away. These findings not only provide a clear methodology to disentangle competing photocurrent mechanisms, but also establish a new paradigm for controlling electrical signals in metals using light. The ability to "write" and reverse current paths on-demand with polarization opens exciting prospects for developing reconfigurable, all-optical nanocircuitry for ultrafast signal processing and on-chip information conversion.

**Acknowledgements**

This work is supported by the ERC grant FemtoMagnet (grant no. 101087709). We acknowledge the financial support from the Agence Nationale de la Recherche (ANR-20-CE09-0031-01, ANR-22-CE09-0027-04 and ANR-23-ERCC-0005), the Institut de Physique du CNRS (Tremplin@INP 2020).

**Author contributions**

A.B. and M.M. conceived of and designed the research. R.Z., D.S., O.M., C.H., A.B. and M.M. carried out the experiments and data collection. R.Z., Y.Y., Y.M. and G.C-d-F. carried out the simulations. R.Z., D.S., O.M., C.H., A.B., X.Y., G.C-d-F., M.S-P., C.S., B.G. and M.M. analyzed the data. M.M. prepared the original draft with contribution of all the authors.

**Competing interests**

The authors declare no competing interests.

**Correspondence and requests for materials** should be addressed to Mathieu Mivelle.

# Bibliography

1. Ma, Q., Krishna Kumar, R., Xu, S.-Y., Koppens, F.H. & Song, J.C. Photocurrent as a multiphysics diagnostic of quantum materials. *Nature Reviews Physics* **5**, 170-184 (2023).
2. Yu, W.J.*, et al.* Unusually efficient photocurrent extraction in monolayer van der Waals heterostructure by tunnelling through discretized barriers. *Nat. Commun.* **7**, 13278 (2016).
3. Zhang, Y.*, et al.* Switchable magnetic bulk photovoltaic effect in the two-dimensional magnet CrI3. *Nat. Commun.* **10**, 3783 (2019).
4. Kastl, C., Karnetzky, C., Karl, H. & Holleitner, A.W. Ultrafast helicity control of surface currents in topological insulators with near-unity fidelity. *Nat. Commun.* **6**, 6617 (2015).
5. Watanabe, H. & Yanase, Y. Chiral photocurrent in parity-violating magnet and enhanced response in topological antiferromagnet. *Physical Review X* **11**, 011001 (2021).
6. Sun, D.*, et al.* Ultrafast hot-carrier-dominated photocurrent in graphene. *Nat. Nanotechnol.* **7**, 114-118 (2012).
7. Son, B.H.*, et al.* Imaging ultrafast carrier transport in nanoscale field-effect transistors. *ACS Nano* **8**, 11361-11368 (2014).
8. Zolotavin, P., Evans, C.I. & Natelson, D. Substantial local variation of the Seebeck coefficient in gold nanowires. *Nanoscale* **9**, 9160-9166 (2017).
9. Zolotavin, P., Evans, C. & Natelson, D. Photothermoelectric effects and large photovoltages in plasmonic Au nanowires with nanogaps. *The Journal of Physical Chemistry Letters* **8**, 1739-1744 (2017).
10. Mennemanteuil, M.-M.*, et al.* Laser-induced thermoelectric effects in electrically biased nanoscale constrictions. *Nanophotonics* **7**, 1917-1927 (2018).
11. Wang, X., Evans, C.I. & Natelson, D. Photothermoelectric detection of gold oxide nonthermal decomposition. *Nano. Lett.* **18**, 6557-6562 (2018).
12. Abbasi, M., Evans, C.I., Chen, L. & Natelson, D. Single metal photodetectors using plasmonically-active asymmetric gold nanostructures. *ACS Nano* **14**, 17535-17542 (2020).
13. Evans, C.I.*, et al.* Thermoelectric response from grain boundaries and lattice distortions in crystalline gold devices. *Proceedings of the National Academy of Sciences* **117**, 23350-23355 (2020).
14. Evans, C.I.*, et al.* Detection of Trace Impurity Gradients in Noble Metals by the Photothermoelectric Effect. *J. Phys. Chem. C* **125**, 17509-17517 (2021).
15. Zhao, B.*, et al.* Photonic Contributions to the Apparent Seebeck Coefficient of Plasmonic Metals. *Nano. Lett.* (2025).
16. Mennemanteuil, M.M., Buret, M., Colas-des-Francs, G. & Bouhelier, A. Optical rectification and thermal currents in optical tunneling gap antennas. *Nanophotonics* **11**, 4197-4208 (2022).
17. Vengurlekar, A.S. & Ishihara, T. Surface plasmon enhanced photon drag in metal films. *Appl. Phys. Lett.* **87**(2005).
18. Noginova, N., Yakim, A., Soimo, J., Gu, L. & Noginov, M. Light-to-current and current-to-light coupling in plasmonic systems. *Physical Review B—Condensed Matter and Materials Physics* **84**, 035447 (2011).
19. Knight, M.W., Sobhani, H., Nordlander, P. & Halas, N.J. Photodetection with active optical antennas. *Science* **332**, 702-704 (2011).
20. Sheldon, M.T., Van de Groep, J., Brown, A.M., Polman, A. & Atwater, H.A. Plasmoelectric potentials in metal nanostructures. *Science* **346**, 828-831 (2014).
21. van de Groep, J., Sheldon, M.T., Atwater, H.A. & Polman, A. Thermodynamic theory of the plasmoelectric effect. *Scientific reports* **6**, 23283 (2016).
22. Cheng, O.H.-C., Son, D.H. & Sheldon, M. Light-induced magnetism in plasmonic gold nanoparticles. *Nat. Photonics* **14**, 365-368 (2020).
23. Hareau, C., Yang, X., Sanz-Paz, M., Sheldon, M. & Mivelle, M. Engineering Magnetization with Photons: Nanoscale Advances in the Inverse Faraday Effect for Metallic and Plasmonic Architectures. *ACS Photonics* (2025).
24. Hertel, R. Theory of the inverse Faraday effect in metals. *J. Magn. Magn. Mater.* **303**, L1-L4 (2006).

25. Hertel, R. & Fähnle, M. Macroscopic drift current in the inverse Faraday effect. *Phys. Rev. B* **91**, 020411 (2015).
26. Nadarajah, A. & Sheldon, M.T. Optoelectronic phenomena in gold metal nanostructures due to the inverse Faraday effect. *Opt. Express* **25**, 12753-12764 (2017).
27. Yang, X.*, et al.* Tesla-Range Femtosecond Pulses of Stationary Magnetic Field, Optically Generated at the Nanoscale in a Plasmonic Antenna. *ACS Nano* **16**, 386-393 (2021).
28. Cheng, O.H.-C., Zhao, B., Brawley, Z., Son, D.H. & Sheldon, M.T. Active Tuning of Plasmon Damping via Light Induced Magnetism. *Nano. Lett.* **22**, 5120-5126 (2022).
29. Yang, X., Hareau, C. & Mivelle, M. Twisting Light, Steering Spins: Gold Nanoparticle Magnetization via Inverse Faraday Effect and Orbital Angular Momentum. *Nano. Lett.* (2025).
30. Yang, X., Mou, Y., Gallas, B., Bidault, S. & Mivelle, M. From dark modes to topology: light-induced skyrmion generation in a plasmonic nanostructure through the inverse faraday effect. *Nanophotonics* (2025).
31. Yang, X.*, et al.* An inverse Faraday effect generated by linearly polarized light through a plasmonic nano-antenna. *Nanophotonics* **12**, 687-694 (2023).
32. Mou, Y., Yang, X., Gallas, B. & Mivelle, M. A chiral inverse Faraday effect mediated by an inversely designed plasmonic antenna. *Nanophotonics* **12**, 2115-2120 (2023).
33. Mou, Y.*, et al.* Femtosecond drift photocurrents generated by an inversely designed plasmonic antenna. *Nano. Lett.* **24**, 7564-7571 (2024).
34. Mou, Y., Yang, X., Gallas, B. & Mivelle, M. A reversed inverse Faraday effect. *Advanced Materials Technologies* **8**, 2300770 (2023).
35. Perrier, F. Université de Lyon (2021).
36. Arbouet, A.*, et al.* Direct measurement of the single-metal-cluster optical absorption. *Phys. Rev. Lett.* **93**, 127401 (2004).
37. Gauchet, M.*, et al.* Bright-Field Polarimetry of a Single Plasmonic Nanostructure Combining Polarization and Position Modulation Techniques. *ACS Photonics* (2025).

# All-optical directional switching of non-thermal photocurrents in plasmonic nanocircuits

Roméo Zapata[1°], Diana Singh[2°], Obren Markovic[1], Chantal Hareau[1], Xingyu Yang[1], Ye Mou[3], Catherine Schwob[1], Bruno Gallas[1], Maria Sanz-Paz[1], Gérard Colas-des-Francs[2], Alexandre Bouhelier[2] and Mathieu Mivelle[1*]

[1]Sorbonne Université, CNRS, Institut des NanoSciences de Paris, INSP, F-75005 Paris, France

[2]Université Bourgogne Europe, CNRS, Laboratoire Interdisciplinaire Carnot de Bourgogne ICB UMR 6303, 21000 Dijon, France

[3]School of Electronic and Information Engineering, Ningbo University of Technology, No. 201, Fenghua Road, Jiangbei District, Ningbo, Zhejiang, China

° Equal contribution

*Corresponding authors:

mathieu.mivelle@sorbonne-universite.fr

**Experimental Setup:** The nano-strips were characterized using a custom-built scanning photocurrent microscopy (SPCM) setup. A 785 nm laser was modulated in either intensity (using an optical chopper at 1 kHz) or polarization (using a photoelastic modulator, PEM, at 50 kHz). The beam was focused onto the sample via a high numerical aperture (NA 1.49) oil-immersion objective. The sample was mounted on a 3-axis piezo stage for raster scanning. The resulting current was amplified by a transimpedance amplifier (TIA, $10^3$ V/A gain) and demodulated using a lock-in amplifier referenced to the optical modulation frequency. This method allows for the separation of the photothermal response (measured via intensity modulation) from the polarization-dependent drift currents (measured via polarization modulation).

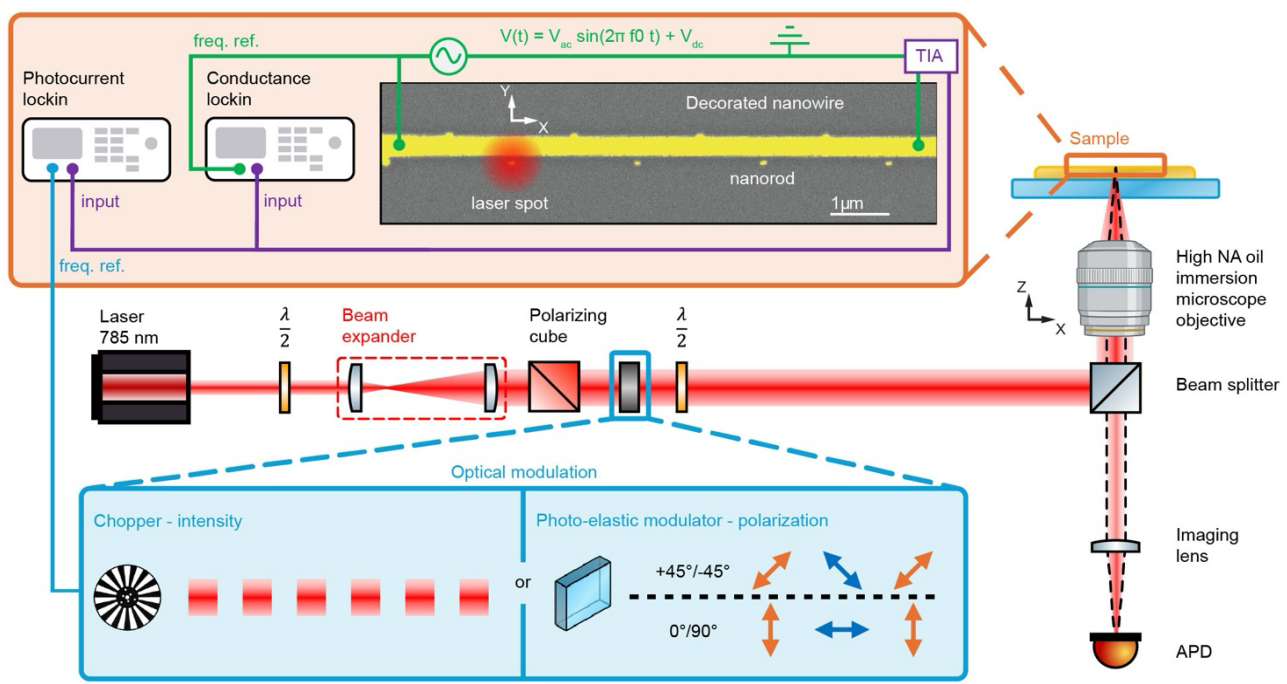

**Figure S1. Schematic of the scanning photocurrent microscopy setup.** The experimental setup is designed to isolate drift photocurrents from photothermal contributions using a dual modulation and detection scheme. The optical excitation is modulated either in intensity using an optical chopper (to measure the photothermal response) or in polarization using a photoelastic modulator (PEM, to isolate the drift current). A half-wave plate sets the final orientation of the polarization relative to the sample. The beam is focused onto the sample by a high numerical aperture (NA = 1.49) oil-immersion objective. The sample is mounted on a three-axis piezo stage for precise positioning and raster scanning. Reflected light is collected by an avalanche photodiode (APD) to create a confocal reflection image. For electrical measurements, an AC voltage ($V_{AC}$) superimposed on a DC offset ($V_{DC}$) is applied across the gold strip, and the current is measured with a transimpedance amplifier (TIA). The TIA signal is then fed to a lock-in amplifier. Demodulation at the optical modulation frequency isolates the photocurrent, while demodulation at the AC bias frequency measures the local conductivity of the strip.

**Data Interpretation and the Dual-Modulation Artifact:**

A critical aspect of our data interpretation is understanding the precise nature of the lock-in signal. In an ideal experiment, polarization modulation between +45° and -45° would yield a signal *S* proportional to the simple difference in photocurrents: $S \propto I(+45°) - I(-45°)$

However, we observed that the PEM introduces a known parasitic effect: a small, periodic positional modulation (i.e., a beam wobble) in addition to the intended polarization modulation. This was confirmed by measuring the beam's deviation on a quadrant photodiode, which revealed a ~ 540 nm wobble at the modulation frequency.

Therefore, the system is subject to a dual modulation. The lock-in amplifier, demodulating at 50 kHz, detects changes originating from both the polarization switch and the parasitic beam movement. As established in the literature [1-3], when a small positional modulation δr is applied simultaneously with a state modulation (like polarization), the resulting lock-in signal is no longer the simple difference Δ*I* but is instead proportional to its spatial derivative.

The demodulation process thus effectively performs two "derivative" operations:

1. A subtraction (the first "derivative") between the two polarization states: $\Delta I = I(+45°) - I(-45°)$.
2. A spatial derivative of this difference, caused by the beam wobble.

The measured signal ($S_{measured}$) is therefore proportional to the spatial derivative of the polarization-difference signal, taken along the wobble axis δr:

$$S_{measured} \propto \nabla(\Delta I) \cdot \delta r = \nabla\big(I(+45°) - I(-45°)\big) \cdot \delta r$$

This explains why our experimental maps (e.g., Fig. 1e) exhibit a characteristic derivative-like (negative-positive-negative) signature. For a rigorous comparison, we therefore compare our experimental data not to the theoretical current maps (Fig. 1g), but to the calculated spatial derivative of these maps (Fig. 1f).

Conversely, to retrieve the "pure" drift current profile from the experimental data, we perform a numerical integration on the measured line cuts (Fig. 1i), which reverses the spatial derivative artifact and confirms the excellent match with theory.

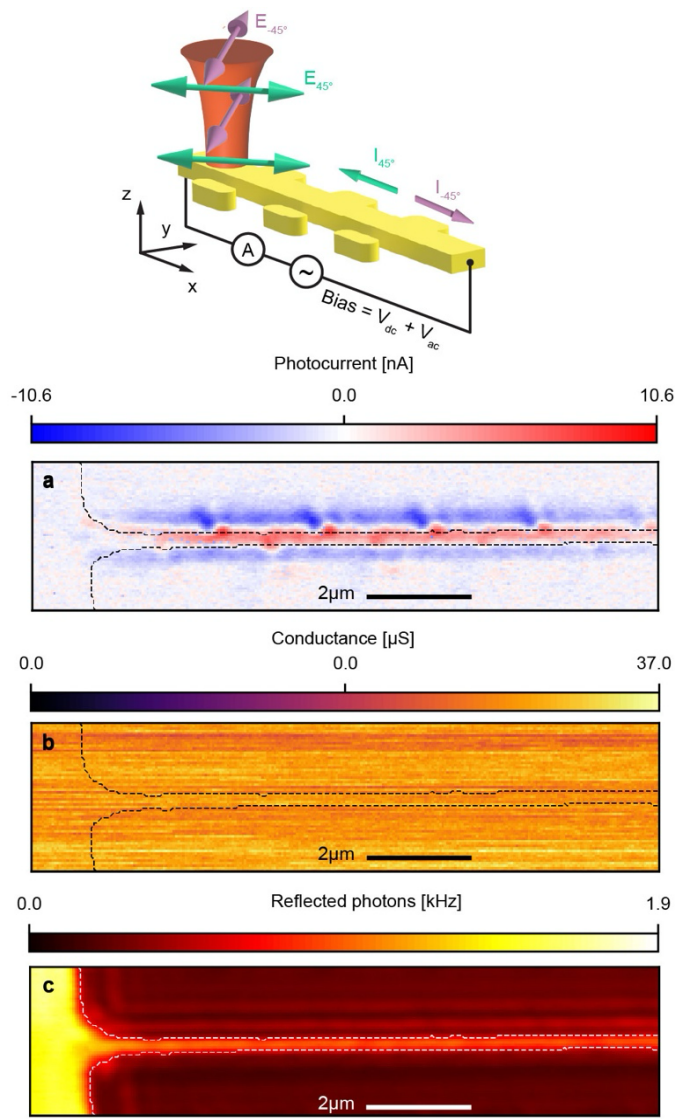

**Figure S2. Parallel mapping of local conductivity and drift photocurrent. a**, Schematic of the dual-modulation setup. An AC bias is applied to the sample simultaneously with a polarization modulation (+45° to -45°) of the optical excitation. **b**, The photocurrent map, obtained by demodulating the signal at the optical modulation frequency, reveals the local generation of polarization-dependent drift currents. **c**, The conductivity map, obtained by demodulating the signal at the AC bias frequency, shows uniform conductivity across the gold strip, independent of the optical excitation. **d**, The confocal reflection map is used to locate the gold strip. An edge-detection algorithm provides the white overlay outline used in all maps. For these measurements, the applied bias consisted of a -50 mV AC component superimposed on a +50 mV DC offset.

**Numerical Modeling of Photothermal Effects**

We numerically investigated the photothermal response using the finite element method (FEM) in COMSOL Multiphysics. The model geometry consists of a gold nanowire (40 nm thick, 290 nm wide, and 10 μm long) connected to two 200 μm electrodes, defining the ground and bias ($V_0$) potentials. Importantly, the Seebeck coefficient of gold is known to depend on film thickness and the width of the electrode or nanowire. Following Ref [4], we modeled this spatial variation by setting the Seebeck coefficient to 1.76 μV/K in the electrodes and 1.82 μV/K in the nanowire.

Based on experimental data, the laser absorption was modeled as a 30 μW heat source deposited in a Gaussian profile with a 500 nm waist. The simulation boundaries included conductive heat transfer within the nanowire and the glass substrate, and convective heat flux at the sample-air interface.

We first analyzed the 0 V bias condition. Figures S6(a-c) show the calculated temperature increase in the nanowire for three different laser positions: -3, 0, and +3 μm. These profiles confirm the presence of opposite temperature gradients at the wire ends, which generate photothermal currents. The resulting estimated photothermal current is plotted as a function of the laser's position in Figure S6(d). As shown, the current varies linearly from a negative to a positive value, passing through zero at the wire's center. This anti-symmetric profile is in good agreement with the opposing temperature gradients and is the characteristic signature of a photothermoelectric response.

We then investigated the finite-bias condition (e.g., +75 mV). In the presence of a bias, the wire's baseline temperature is dominated by Joule heating, increasing by 30°C at the center (Figures S6(e-g)). The total circuit current is a sum of this large Ohmic current and the small photothermal contribution. As shown in Figure S6(h), the current now flows in the same (positive) direction for all laser positions, but it retains a photothermal contribution (slope) similar to that in Figure S6(d), driven by the same thermal gradients. These results are in perfect agreement with the experimental data presented in Figs. 1c,d and 3 of the main manuscript.

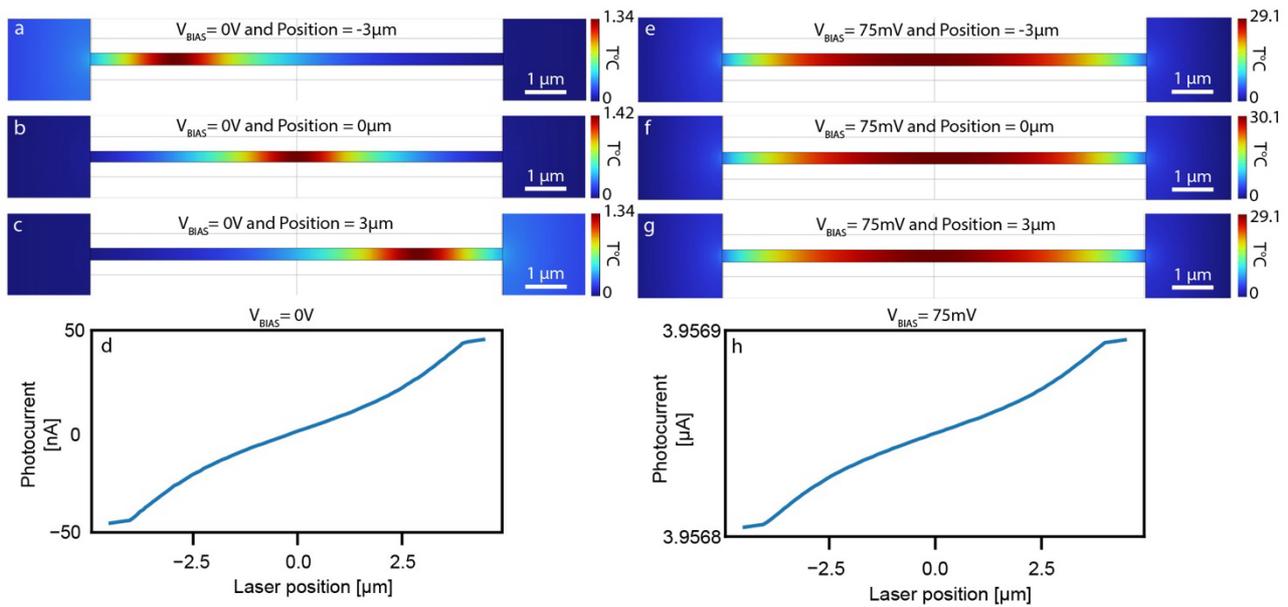

**Figure S3. FEM modeling of the photothermal response in the gold nanowire. a-c**, Simulated temperature increase for three laser positions (at -3 μm, 0 μm, and +3 μm from the center, respectively) with 0 V bias, showing opposite thermal gradients at the wire ends. **d**, The resulting estimated photothermal current vs. laser position (0 V bias). The current varies linearly from negative to positive, crossing zero at the center. This anti-symmetric profile is the characteristic signature of a photothermoelectric response, agreeing with the opposing gradients seen in (**a-c**). **e-g**, Simulated temperature increase for the same three laser positions at +75 mV bias. The wire's baseline temperature is dominated by a 30°C increase from Joule heating. **h**, The total circuit current vs. laser position (+75 mV bias). The current is now positive for all positions (dominated by the Ohmic current) but retains a photothermal contribution (slope) similar to that in (**d**). These results are in perfect agreement with the experimental thermal responses shown in Figs. 1c,d and 3.

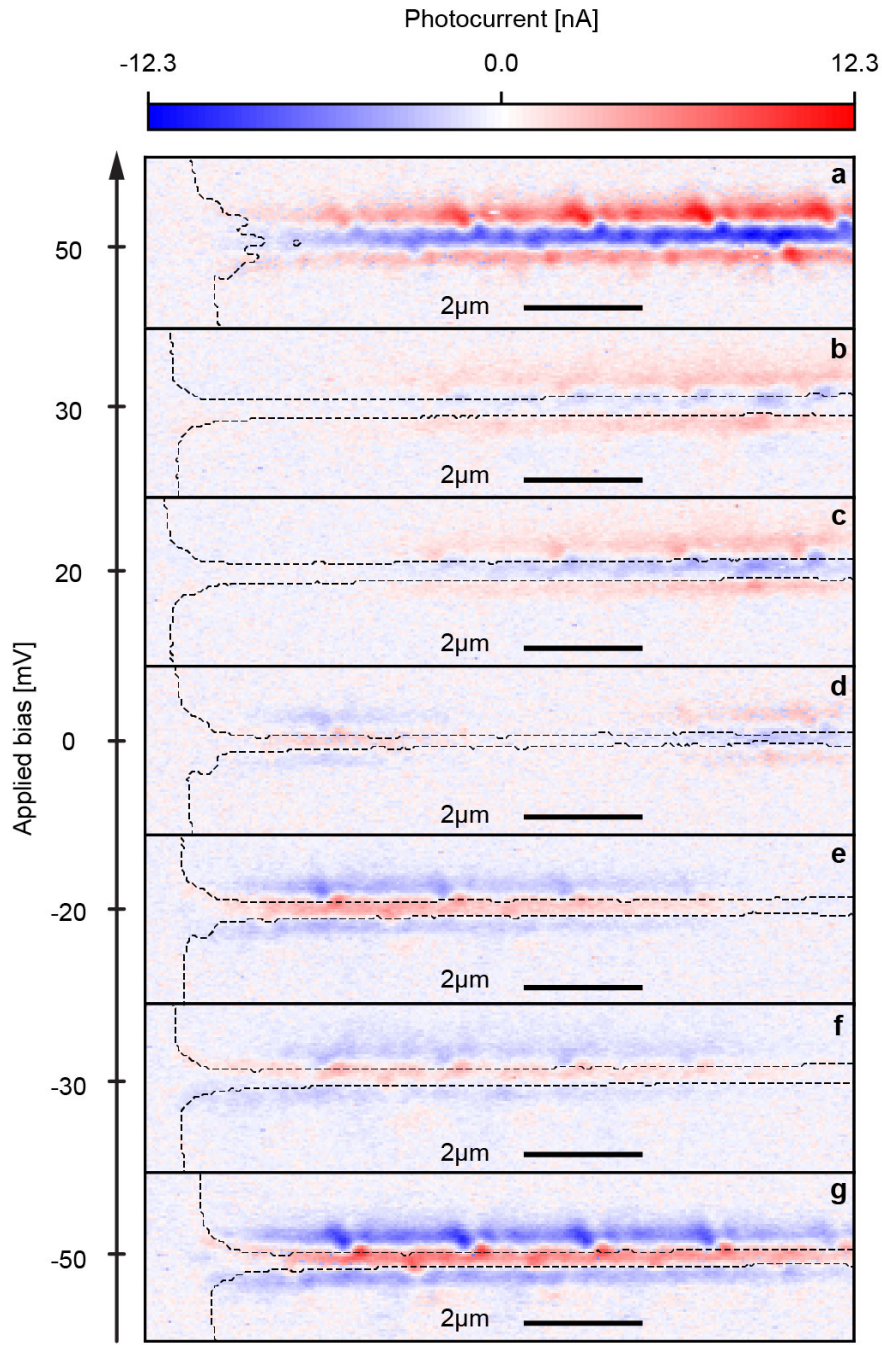

**Figure S4. Dependence of drift photocurrent intensity on applied DC bias. a–g**, A series of experimental photocurrent maps of the both-side nanorod-decorated gold strip, recorded under polarization modulation (+45° to -45°) for seven different DC bias voltages, ranging from +50 mV to -50 mV. The absolute intensity of the drift photocurrent signal increases significantly as the magnitude of the applied bias increases. A clear sign inversion in the photocurrent map is observed as the bias crosses 0 mV. This demonstrates that while the generation of the drift current is an all-optical process, the efficiency of its collection is strongly modulated by the applied DC field.

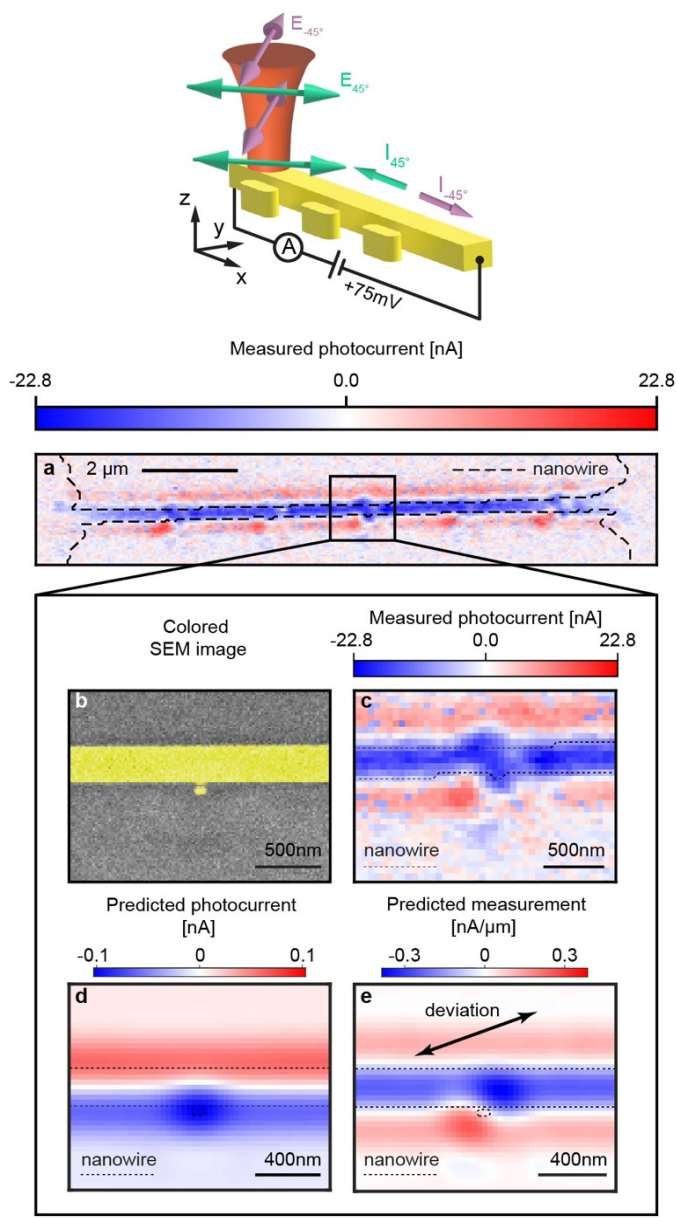

**Figure S5. Drift photocurrent control using capacitively coupled nanorods.** a, Experimental photocurrent map of a gold strip decorated only on one side with nanorods separated by a nominal 20 nm gap. The map, recorded under polarization modulation, shows a transverse sign inversion overlaid with periodic longitudinal features. b, Scanning electron microscope (SEM) image of a single detached nanorod. c, Magnified view of the experimental photocurrent map from the region near a detached nanorod. d, Theoretical map of the pure drift current distribution in the same region, based on the Inverse Faraday Effect. e, The spatial derivative of the theoretical map in (d), calculated along the axis of the PEM-induced beam wobble (arrow). This map serves as the direct theoretical counterpart to the experimental measurement in (c) and shows excellent agreement.

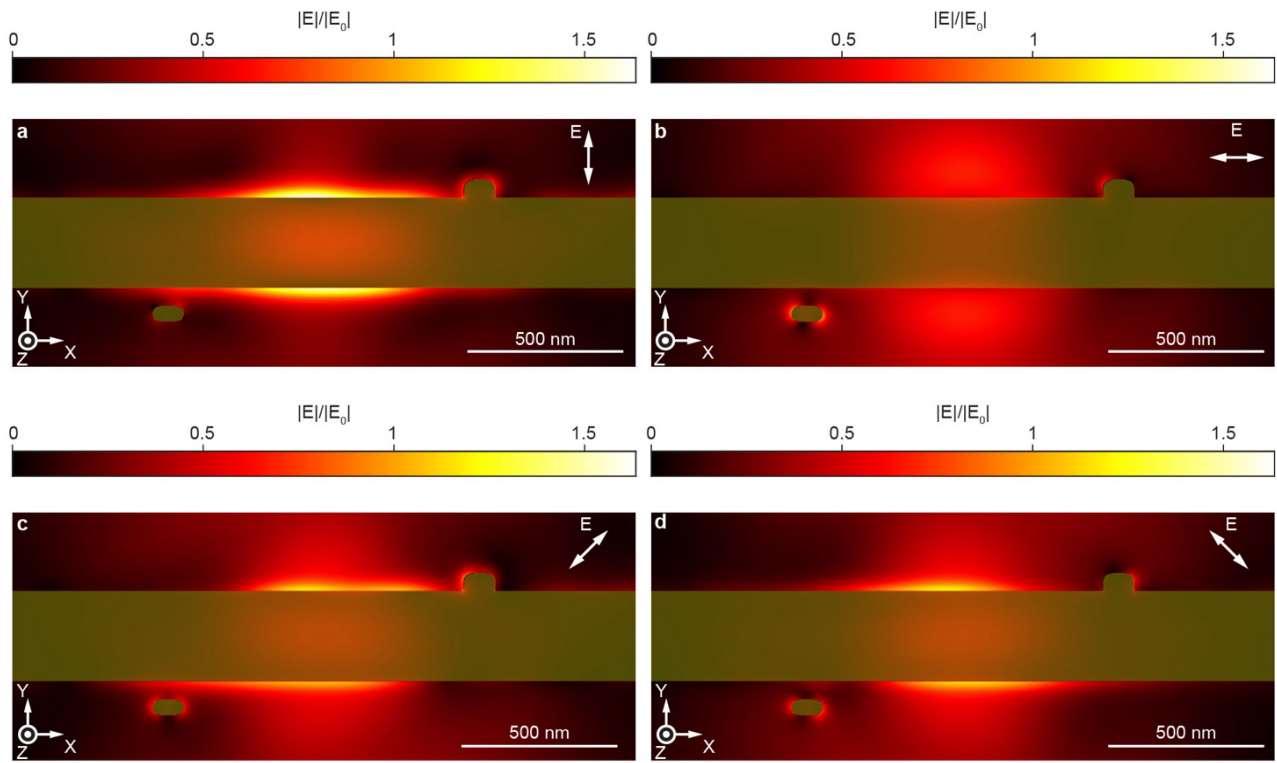

**Figure S6. Simulated near-field enhancement for different linear polarizations.** Simulated maps of the plasmonic near-field enhancement ($|E/E_0|^2$) around a both-side nanorod-decorated gold strip for four different incident linear polarizations. The polarization is oriented at **a**, 90° (perpendicular), **b**, 0° (parallel), **c**, +45°, and **d**, -45° with respect to the strip's main axis. The field enhancement is maximized for the perpendicular polarization, exciting a strong plasmonic resonance, and minimized for the parallel polarization. The enhancement for +45° and -45° polarizations is symetrically identical, which is the key condition that allows for the isolation of non-thermal drift currents from the thermal background in our experiments. The nanostructure's geometry is indicated by the colored overlay.

[1] A. Arbouet et al, "Direct Measurement of the Single-Metal-Cluster Optical Absorption" *Phys. Rev. Lett.* 93, 127401, 2004

[2] F. Perrier, "Etude expérimentale de la réponse optique de nanostructures métalliques par modulation de polarisation: application à la chiro-plasmonique," Université de Lyon (2021).

[3] M. Gauchet et al., "Bright-Field Polarimetry of a Single Plasmonic Nanostructure Combining Polarization and Position Modulation Techniques," *ACS Photonics*, vol. pp. 2025.

[4] P. Zolotavin et al. "Photothermoelectric Effects and Large Photovoltages in Plasmonic Au Nanowires with Nanogaps", *Phys. Chem. Lett.* 8, 1739–1744 (2017).